\documentclass[a4paper,11pt]{article}
\usepackage{pos}

\usepackage{wrapfig}
\usepackage{subcaption}
\usepackage{siunitx}

\renewcommand{\logo}{\relax}

\title{The IDEA detector concept for FCC-ee}

\author*{Armin Ilg}

\affiliation{Physik Institut, University of Zürich\\
Winterthurerstrasse 190, 8057 Zürich, Switzerland}

\onbehalf{on behalf of the IDEA Study Group and FCC}

\emailAdd{armin.ilg@cern.ch}

\abstract{
    The electron-positron stage of the Future Circular Collider (FCC-ee) provides exciting opportunities that are enabled by next generation particle physics detectors. This contribution presents IDEA, a detector concept optimised for FCC-ee and composed of a vertex detector based on MAPS, a very light drift chamber, a silicon wrapper, a high resolution dual-readout crystal electromagnetic calorimeter, an HTS based superconducting solenoid, a dual-readout fibre calorimeter, and three layers of muon chambers embedded in the magnet flux return yoke. In particular, the physics requirements and the technical solutions chosen in the various sub-systems to address them are discussed. This is followed by a description of the detector R\&D currently in progress, test-beam results, and the expected performance on some key physics benchmarks.
}

\FullConference{The European Physical Society Conference on High Energy Physics (EPS-HEP2025)\\
7-11 July 2025\\
Marseille, France\\}

\begin{document}
\maketitle

\section{Introduction}

The Future Circular Collider (FCC \cite{FCC_FS_Exp,FCC_FS_Acc,FCC_FS_Impl_Sust}) is a proposed facility to succeed the LHC as the next large collider facility, serving the particle physics community throughout the 21st century. In a first stage, the \textit{FCC-ee} would collide electrons and positrons at centre-of-mass ($\sqrt{s}$) energies from \SI{88}{GeV} (just below the Z pole) up to \SI{365}{\giga\electronvolt} (above the $t\bar{t}$ threshold), providing unprecedented luminosities and thus tiny statistical uncertainties. The particle physics community currently explores various detector concepts to instrument the four interaction points of FCC-ee.

\begin{wrapfigure}[13]{r}{0.5\textwidth}
    \centering
    \vspace*{-0.45cm}
    \includegraphics[width=\linewidth,trim=0cm 0.1cm 0cm 0cm,clip]{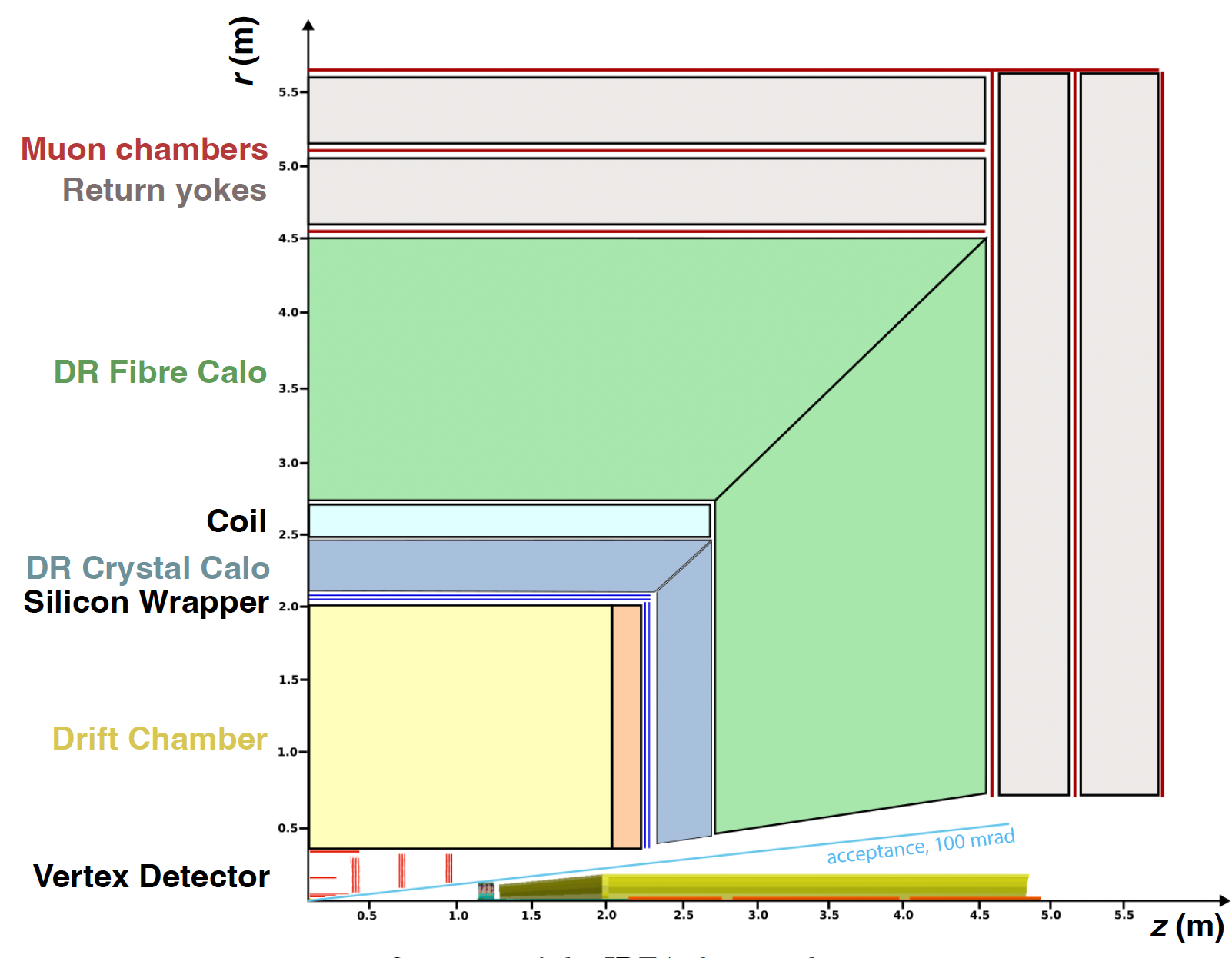}
    \caption{IDEA detector concept layout~\cite{IDEAStudyGroup:2025gbt}.}
    \label{fig:IDEA_layout}
\end{wrapfigure}

The IDEA detector concept~\cite{IDEAStudyGroup:2025gbt} adopts numerous novel instrumentation technologies, targeting to fulfil the \textit{aggressive} subdetector requirements of the FCC-ee\footnote{Table 13 in Reference~\cite{FCC_FS_Exp}.}. The layout of the IDEA detector concept is shown in Figure~\ref{fig:IDEA_layout}, starting with a light-weight central beam pipe design ($\SI{0.65}{\percent}$ of a radiation length $X_0$) and luminosity calorimeter at $z = \SI{1.074}{\meter}$, common to all detector concepts. The tracking system is composed of a silicon vertex detector, a drift chamber, and an outer silicon wrapper. The vertex detector is based on monolithic active pixel sensors (MAPS) that combine signal generation, amplification, and readout into a single silicon die. The very light drift chamber begins at \SI{35}{\centi\meter} of radius and is crucial for precision tracking and particle identification through cluster counting ($\text{d}N/\text{d}x$). The silicon wrapper provides a last precise track hit at about \SI{2}{m} in $r$ and $z$. Both the electromagnetic (ECAL) and hadronic calorimeters (HCAL) follow the dual-readout approach to measure both the hadronic and electromagnetic shower components simultaneously, allowing to reach small energy resolutions. The ECAL is based on crystals, while the HCAL uses fibres. A high-temperature superconducting (HTS) solenoid magnet is placed in-between the ECAL and HCAL, minimising the material in front of the ECAL. The IDEA detector concept is completed by a \textmu-RWELL muon detector in the magnet flux return yokes.

The following sections describe the ongoing detector R\&D and simulation studies of the various IDEA subsystems and discuss the impact on relevant physics benchmarks.








\section{Vertex detector}

The baseline IDEA vertex detector~\cite{Ilg2024} design is based on MAPS available already today. The inner vertex detector uses overlapping staves of ARCADIA~\cite{Pancheri2020}, while the outer vertex detector foresees quad modules of ATLASPix3~\cite{Peric2021} MAPS. It is fully integrated into the FCC-ee machine-detector interface~\cite{Boscolo2025}. The inner vertex detector consists of three single-hit barrel layers, starting at an inner radius of $r_\text{min} = \SI{13.7}{mm}$. The outer vertex consists of two barrel layers at 13 and \SI{31.5}{\centi\meter} radius and three disks per side, providing a good momentum resolution of low to medium momentum tracks and gives a precise hit just before the following drift chamber. 

Assuming $3\times \SI{3}{\micro\meter\squared}$ single-hit resolution for the inner\footnote{To be on equal footing with the CLD vertex detector assuming $3\times \SI{3}{\micro\meter\squared}$ in the first three layers~\cite{CLD_2}.} and $14 \times \SI{43}{\micro\meter\squared}$ for the outer vertex, full simulation of this vertex detector inserted into the CLD~\cite{CLD_2} tracker is performed. Using conformal tracking~\cite{Brondolin2020}, an impact parameter resolution in transverse direction $\sigma(d_0)$ of $\sim \SI{2}{\micro\meter} \oplus \SI{14}{\micro\meter\per\giga\electronvolt}/(p\sin^{3/2}\theta)$ is achieved, surpassing the FCC-ee requirements~\cite{FCC_FS_Exp} already. The important decay channel of $B^0 \rightarrow K^{*0} \tau^+\tau^-$, however, would require an about forty percent improved impact parameter resolution to discover its branching ratio down to the Standard Model prediction~\cite{Miralles:2024Ks}. This motivates further improvements of the beam pipe and vertex detector designs beyond the current baseline, specifically through reduction of their material budget.

\begin{figure}[htbp]
    \centering
    \begin{subfigure}[t]{0.49\textwidth}
        \centering
        \includegraphics[height=0.24\textheight]{"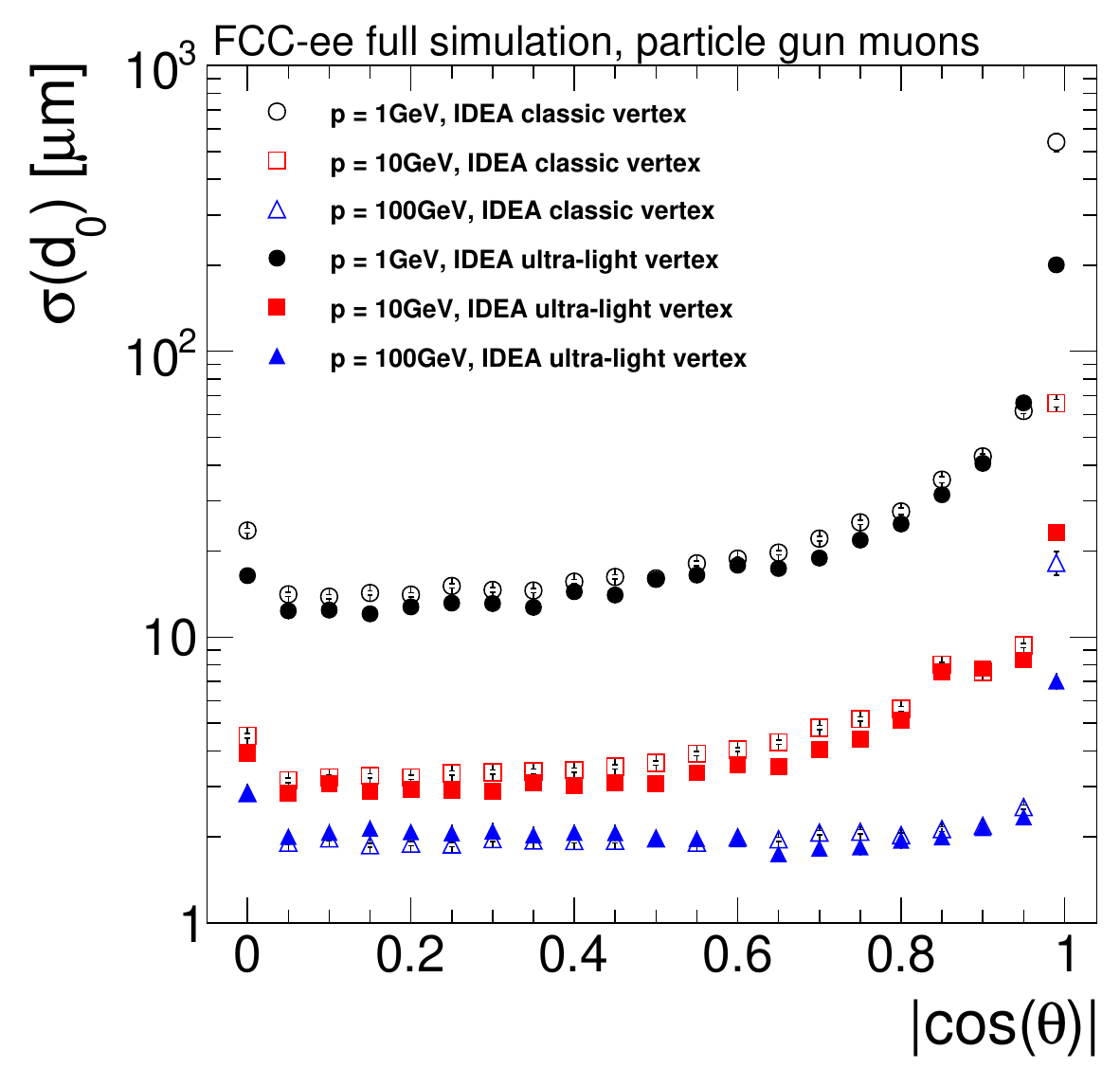"}
        \caption{Impact of ultra-light inner vertex.}
        \label{fig:vertex_comparison_mat}
    \end{subfigure}
    \hfill
    \begin{subfigure}[t]{0.49\textwidth}
        \centering
        \includegraphics[height=0.24\textheight]{"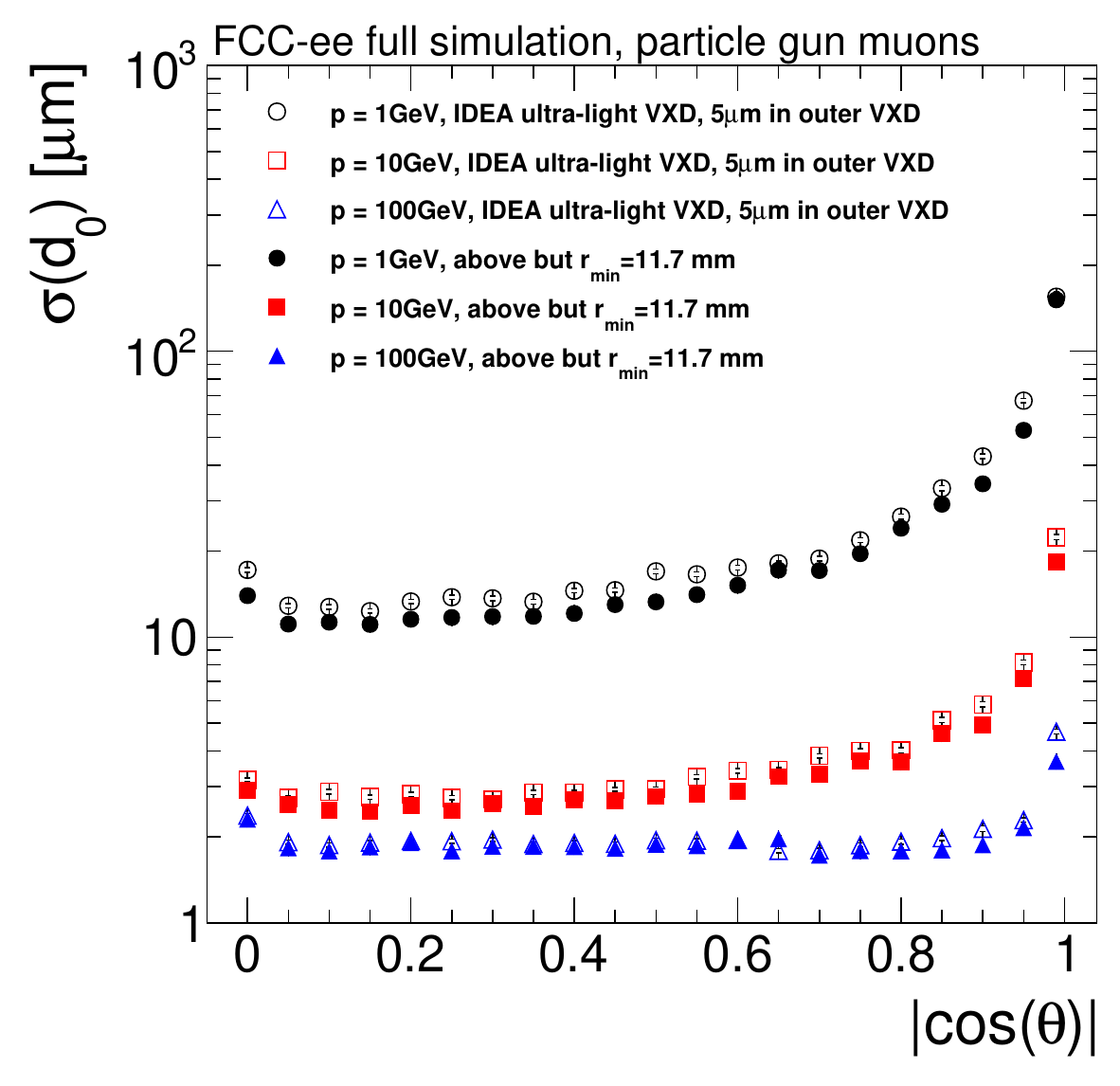"}
        \caption{Impact of improved outer vertex resolution and reduced inner radius for the ultra-light inner vertex.}
        \label{fig:vertex_comparison_betterRes_rmin}
    \end{subfigure}
    \caption{Transverse impact parameter resolution for different configurations of the IDEA vertex detector.}
    \label{fig:vertex_comparison}
\end{figure}

The ultra-light inner vertex detector concept of IDEA~\cite{Ilg2024} adopts an ALICE ITS3-like~\cite{alice_its3_tdr} design using curved, wafer-scale sensors. Four layers are used to help dealing with gaps in detector coverage and the third and fourth layers deploy two sensors per half-barrel to extend the forward coverage. Figure~\ref{fig:vertex_comparison_mat} shows the performance compared to the classic, stave-based design. A $\sigma(d_0)$ resolution of about $\SI{2}{\micro\meter} \oplus \SI{12.5}{\micro\meter\per\giga\electronvolt}/(p\sin^{3/2}\theta)$ is achieved, marking a significant improvement in the low-momentum regime.

Further improvements in the vertexing capabilities can be achieved by improving the spatial resolution of the outer vertex layers and by attaching the first inner vertex barrel layer directly to the beam pipe, reducing its inner radius to \SI{11.7}{mm}. The latter is motivated by the ongoing R\&D on air-cooling of the inner vertex detector. Cooling the first layer ($\approx \SI{12}{\watt}$~\cite{Boscolo2025}) using the beam pipe paraffin cooling would reduce the amount of air-cooling piping. Figure~\ref{fig:vertex_comparison_betterRes_rmin} shows the impact of these improvements on top of the ultra-light design. The improved outer vertex sensor resolution enhances the $d_0$ resolution for \SI{10}{\giga\electronvolt} muons, while the reduced inner radius benefits all displayed momenta. Overall, $\sigma(d_0) \sim \SI{1.8}{\micro\meter} \oplus \SI{11.5}{\micro\meter\per\giga\electronvolt}/(p\sin^{3/2}\theta)$ could be achievable.


\section{Drift chamber}

The IDEA drift chamber is made up of 112 hyperboloidal layers of field and sense wires, each providing a resolution of $\sigma_{xy} \approx \SI{100}{\micro\meter}$. The chamber is filled with a gas mixture of \SI{90}{\percent} helium and \SI{10}{\percent} H$_4$C$_{10}$. Using light-weight mechanics, a material budget of 1.6--\SI{5.0}{\percent} of $X_0$ is achieved, depending on the track polar angle. 

Using fast simulation ML-based tracking, a tracking efficiency greater than $\SI{95}{\percent}$ is achieved already for tracks with $p_T > \SI{100}{\mega\electronvolt}$~\cite{garcia2025geometric}. Besides tracking, the drift chamber is crucial also for particle identification through cluster counting. A fast simulation study has shown that, if paired with a time-of-flight measurement with $\sigma(t) = \SI{100}{\pico\second}$ at $\mathcal{O}(\SI{2}{\meter}$), a kaon-pion separation of greater than $3\sigma$ can be achieved up to \SI{30}{\giga\electronvolt}~\cite{IDEAStudyGroup:2025gbt}. The large number of tracker hits is also useful for long-lived particle searches and the reconstruction of $K_\text{s}$~\cite{Aleksan2021}. Both of these features are crucial for strange tagging at the Z or HZ pole~\cite{Blekman2025}.

\begin{figure}[htbp]
    \centering
    \begin{subfigure}[c]{0.35\textwidth}
        \centering
        \includegraphics[width=\linewidth]{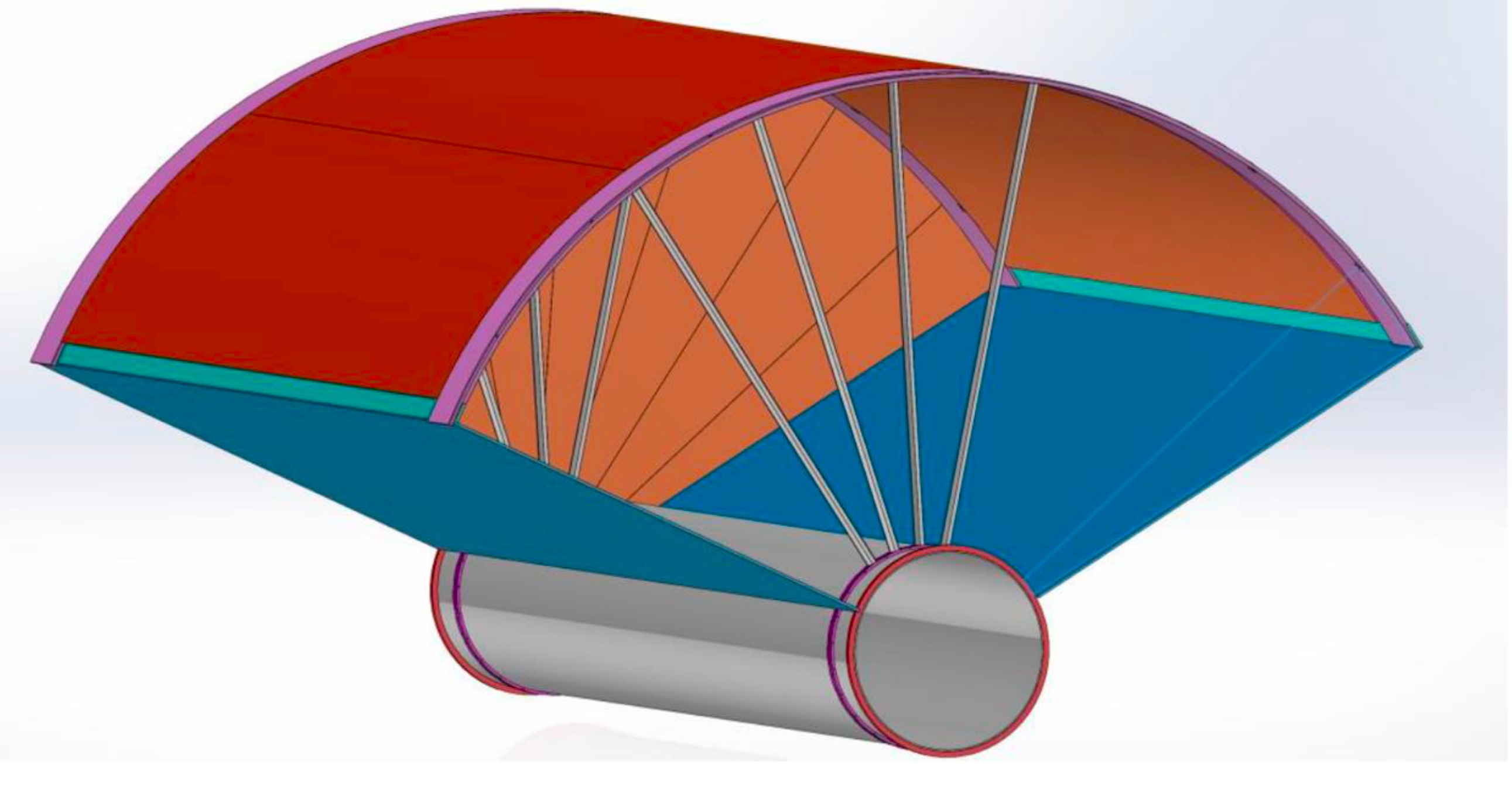}
        \caption{Global layout}
        \label{fig:drift_chamber_proto}
    \end{subfigure}
    \begin{subfigure}[c]{0.64\textwidth}
        \centering
        \includegraphics[width=\linewidth]{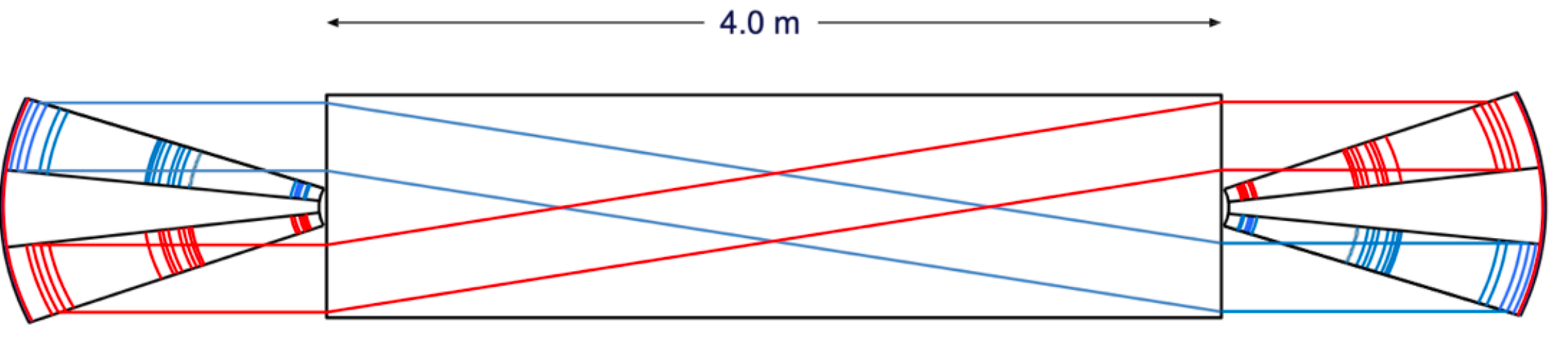}
        \vspace*{0.07cm}
        \caption{Wiring scheme}
        \label{fig:drift_chamber_layout}
    \end{subfigure}
    \caption{Drift chamber prototype under construction, representing one third of the final design.}
    \label{fig:drift_chamber}
\end{figure}

Figure~\ref{fig:drift_chamber} shows the layout of the full-size drift chamber prototype which is currently being constructed to test the mechanical and electrostatic stability. It will represent one third of the drift chamber and feature 1400 wires, compared to 350k wires in the final drift chamber.



\section{Silicon wrapper}\label{sec:silicon_wrapper}

The silicon wrapper follows just outside the drift chamber. With a foreseen spatial resolution of at least $7 \times \SI{90}{\micro\meter\squared}$, it will provide a last, precise track hit. This outer fix point will help calibrating the drift chamber and will provide a stable ruler for the detector acceptance definition. The silicon wrapper also improves the momentum resolution in the high-$p_\text{T}$ regime.

\begin{wrapfigure}[12]{r}{0.4\textwidth}
    \centering
    \vspace*{-0.6cm}
    \includegraphics[width=0.92\linewidth,trim=0.7cm 0.4cm 0.9cm 0.5cm,clip]{"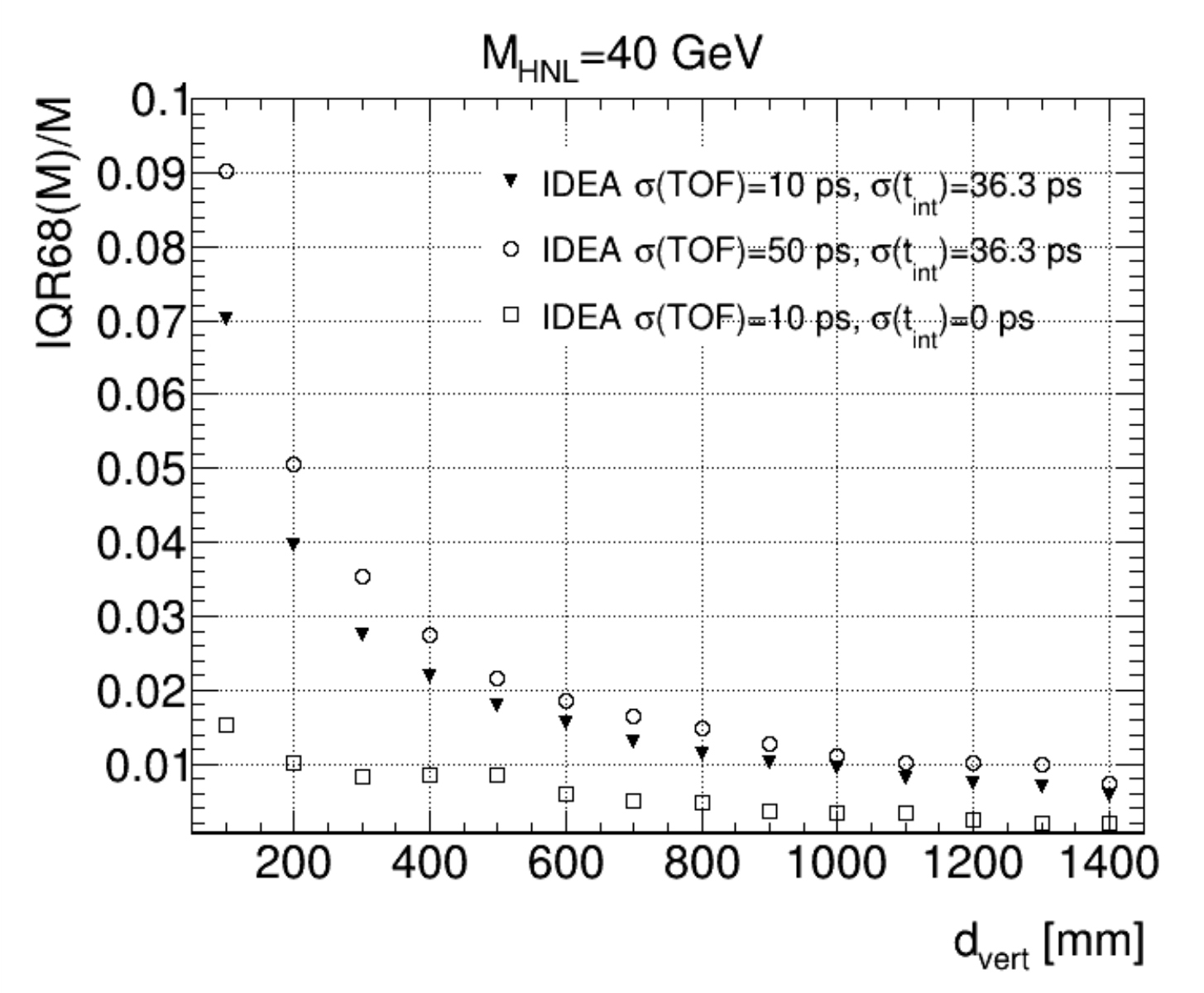"}
    \caption{Mass resolution of a heavy-neutral lepton under different timing performance assumptions~\cite{Aleksan2025}.
    }
    \label{fig:silicon_wrapper}
\end{wrapfigure}

LGADs and MAPS are being explored to potentially equip the silicon wrapper with timing capabilities of $\mathcal{O}(\SI{100}{\pico\second})$. This would complement cluster counting and enable good kaon-pion separation for most momenta. This would also help for long-lived particle searches, though at some point the limiting factor becomes the uncertainty on the timing of the primary interaction. Having a second timing layer, e.g. integrated in the outermost vertex detector layer, could thus improve the measurement of some long-lived particles as illustrated in Figure~\ref{fig:silicon_wrapper}~\cite{Aleksan2025}.  

The silicon wrapper covers an area of about \SI{85}{\meter\squared}. The next iteration of IDEA silicon wrapper in DD4hep~\cite{Gaede2020} will implement a design with only one barrel layer and one disk per side, targeting one hit per track to minimise the sensor area needed. First support concepts are developed as well, foreseeing long carbon fibre longeron and ring structures supporting the silicon wrapper on top of the drift chamber outer wall. The disks will be attached to the drift chamber using stilt-like structures. In both cases, the sensors should face the IP whenever possible.

The next step in the simulation of the complete IDEA tracking system is the track refitting in full simulation, allowing a realistic evaluation of the full tracking system performance, including an updated evaluation of the vertexing performance.

\section{Dual-readout crystal electromagnetic calorimeter}

\begin{figure}[htbp]
    \centering
    \begin{minipage}[c]{0.58\textwidth}
        \centering
        \includegraphics[height=0.20\textheight]{"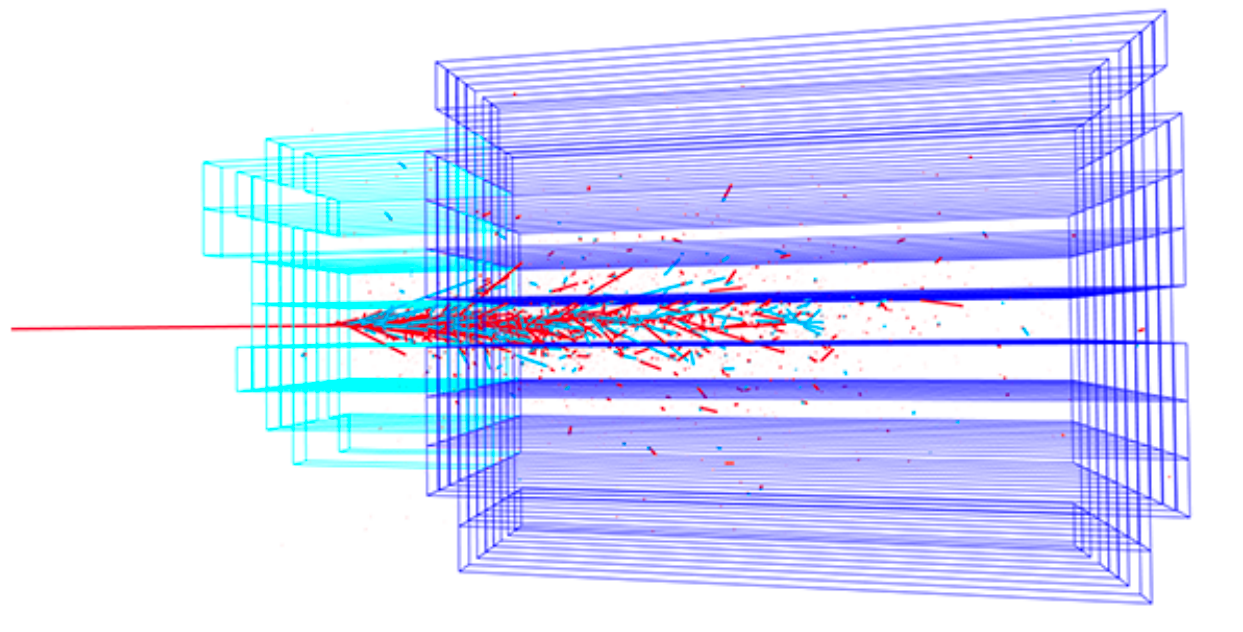"}
        \caption{Simulated electromagnetic shower of a \SI{10}{\giga\electronvolt} electron in the crystal ECAL. Adapted from Reference~\cite{IDEAStudyGroup:2025gbt}.}
        \label{fig:ecal_shower_geant4}
    \end{minipage}
    \hfill
    \begin{minipage}[c]{0.41\textwidth}
        \centering
        \includegraphics[height=0.20\textheight,trim=0cm 0cm 0cm 1.5cm,clip]{"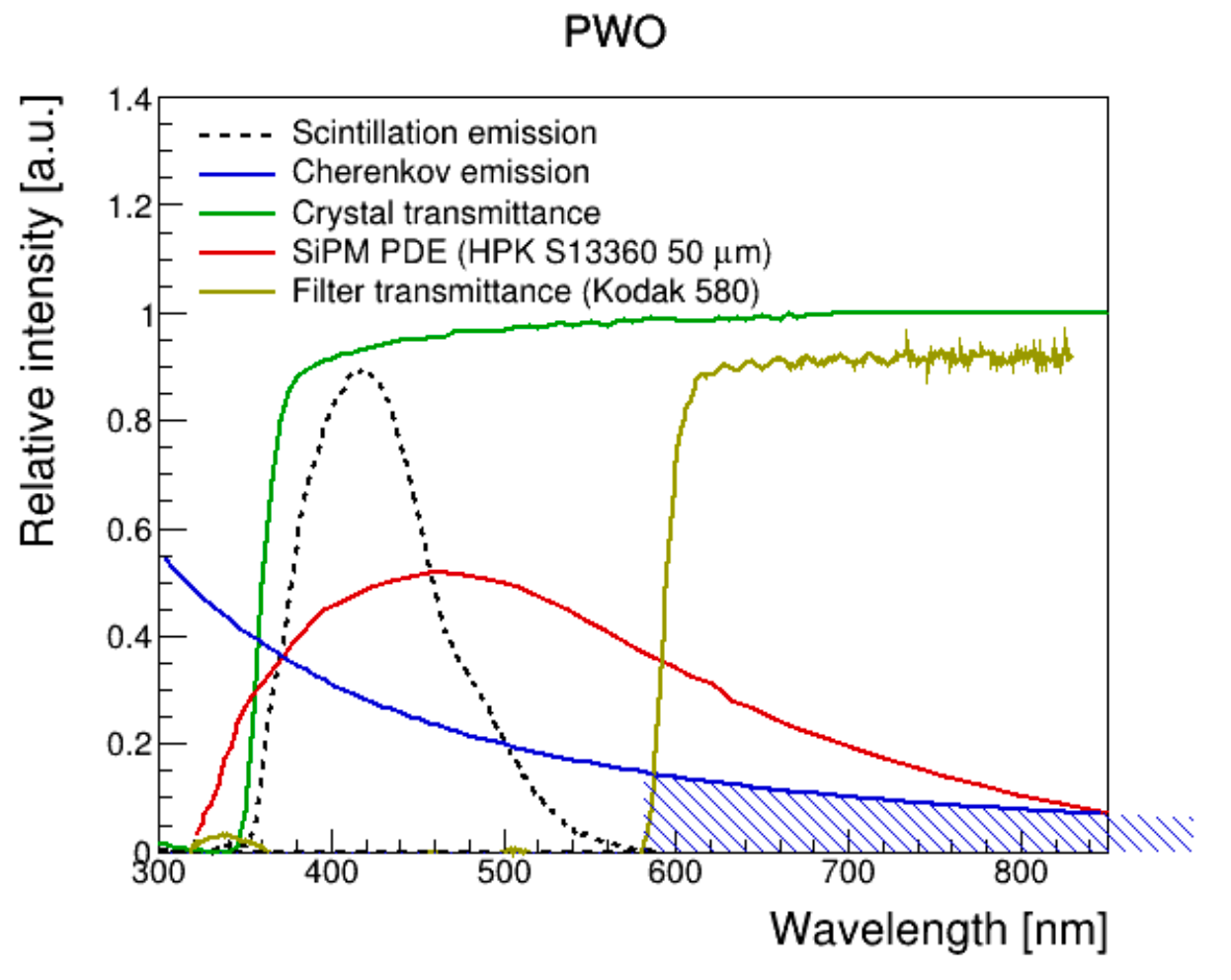"}
        \caption{Dual-readout strategy in PbWO$_4$ crystals. Adapted from Reference~\cite{IDEAStudyGroup:2025gbt}.}
        \label{fig:ecal_PWO_principle}
    \end{minipage}
\end{figure}

The IDEA dual-readout crystal ECAL foresees longitudinally segmented crystals, read out by silicon photomultipliers (SiPMs). The front segment ($6 X_0$) contains a $5 \times \SI{5}{\milli\meter\squared}$ SiPM per crystal, which is optimised for scintillation light detection. The longer, rear crystal segment ($16 X_0$) features two $4\times \SI{4}{mm\squared}$ SiPMs with filters for scintillation and Cherenkov light respectively. A Geant4 study has shown that a total resolution of $\sigma_E/E = \SI{3}{\percent}/\sqrt{E} \oplus \SI{0.5}{\percent}$ is achievable~\cite{Lucchini2020}. Such an ECAL resolution is important at FCC-ee to resolve e.g. $B^0 \rightarrow D_s K$ from $B_s \rightarrow D_s K$~\cite{Aleksan:2021gii,FCC_FS_Exp}. Figure~\ref{fig:ecal_shower_geant4} shows a Geant4 simulation of an electromagnetic shower in the crystal ECAL.

The current baseline choice for the crystal material is PbWO$_4$. Figure~\ref{fig:ecal_PWO_principle} shows the dual-readout strategy of the crystal ECAL based on filtering out scintillation photons using a filter. Other crystal materials under study are BGO and BSO. They feature a peak emission at \SIrange{470}{480}{nm}, leading to a higher light yield but making it harder to filter out scintillation photons. The discrimination between scintillation and Cherenkov light is instead using timing, benefitting from the slower decay time of scintillation light in BGO/BSO.

Recent test beam studies have measured the scintillation and Cherenkov light yields in single PbWO$_4$ crystals as a function of the energy, as shown in Figure~\ref{fig:ecal_PWO_principle}. A $9\times 9$ ECAL prototype is currently being assembled and will be tested at CERN SPS in Fall 2025. 

Lastly, two layers of fast scintillating LYSO:Ce crystals could be added in front of the ECAL to add MIP timing capabilities of $\mathcal{O}(\SI{20}{\pico\second})$, providing similar benefits as discussed in Section~\ref{sec:silicon_wrapper}.

\section{High-temperature superconducting solenoid}

\begin{wrapfigure}[13]{r}{0.34\textwidth}
    \centering
    \vspace*{-0.15cm}
    \includegraphics[width=\linewidth]{"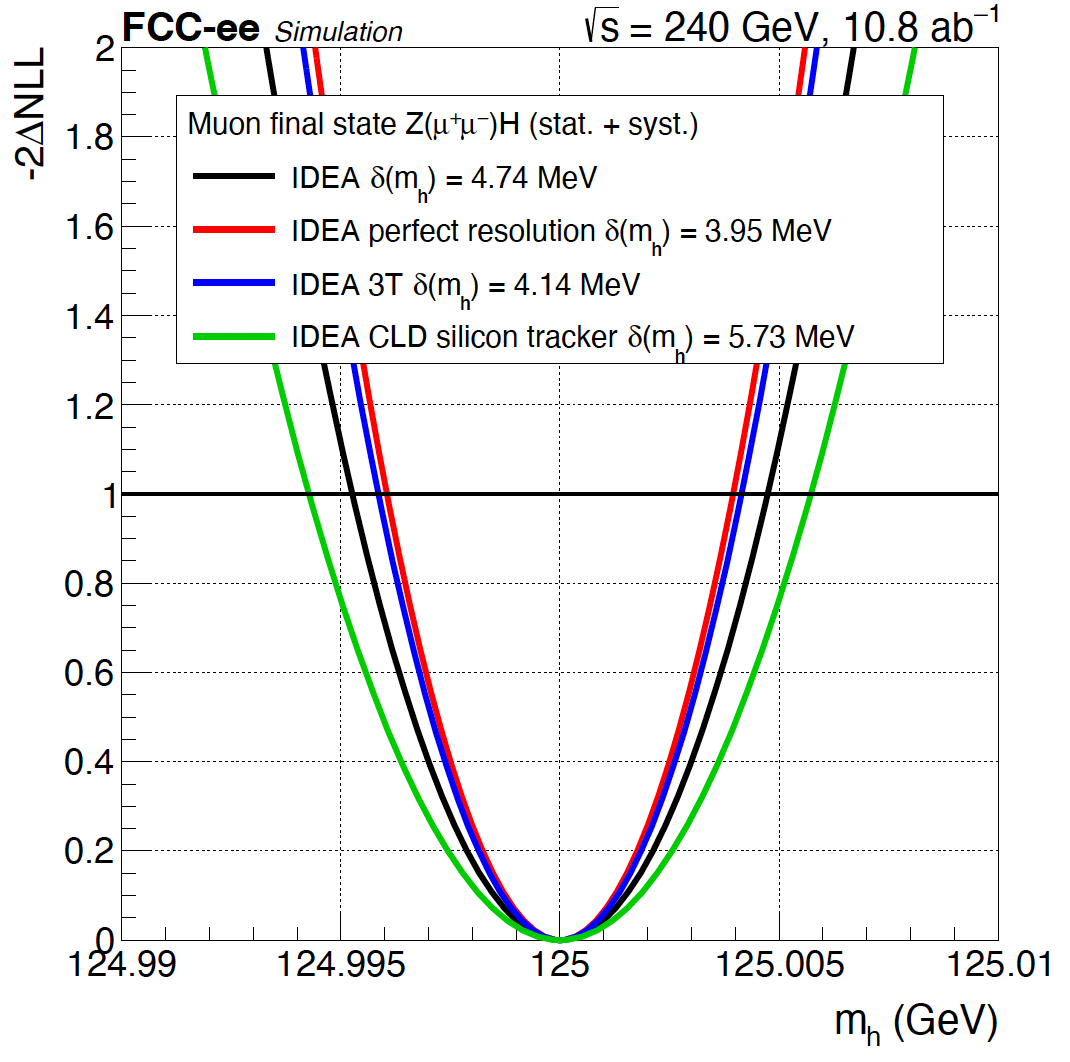"}
    \caption{Higgs mass measurement in $H \rightarrow \mu^+ \mu^-$ for different IDEA detector configurations~\cite{FCC_FS_Exp}.}
    \label{fig:hts_solenoid}
\end{wrapfigure}

In the latest iteration of the IDEA detector concept, the solenoid magnet is moved after the ECAL to fully benefit from the excellent ECAL resolution. With the material budget of the solenoid being less important in such an arrangement, a high-temperature superconducting (HTS) solenoid is foreseen, which could significantly reduce the power consumption and liquid helium inventory needed since an operation temperature of \SI{20}{K} is foreseen, with up to \SI{50}{K} to be considered. 

Currently, in order not to disturb the beam optics, the magnetic field has to be limited to \SI{2}{\tesla} at the Z pole, while \SI{3}{\tesla} are possible at higher $\sqrt{s}$. The HTS solenoid is thus designed to deliver a maximal field of \SI{3}{\tesla}. This would improve the measurement of the Higgs mass by \SI{13}{\percent}
 as shown in Figure~\ref{fig:hts_solenoid}~\cite{FCC_FS_Exp}.

\section{Dual-readout fibre hadronic calorimeter}

\begin{figure}[htbp]
    \centering
    \begin{minipage}[b]{0.54\textwidth}
        \centering
        \includegraphics[height=0.18\textheight]{"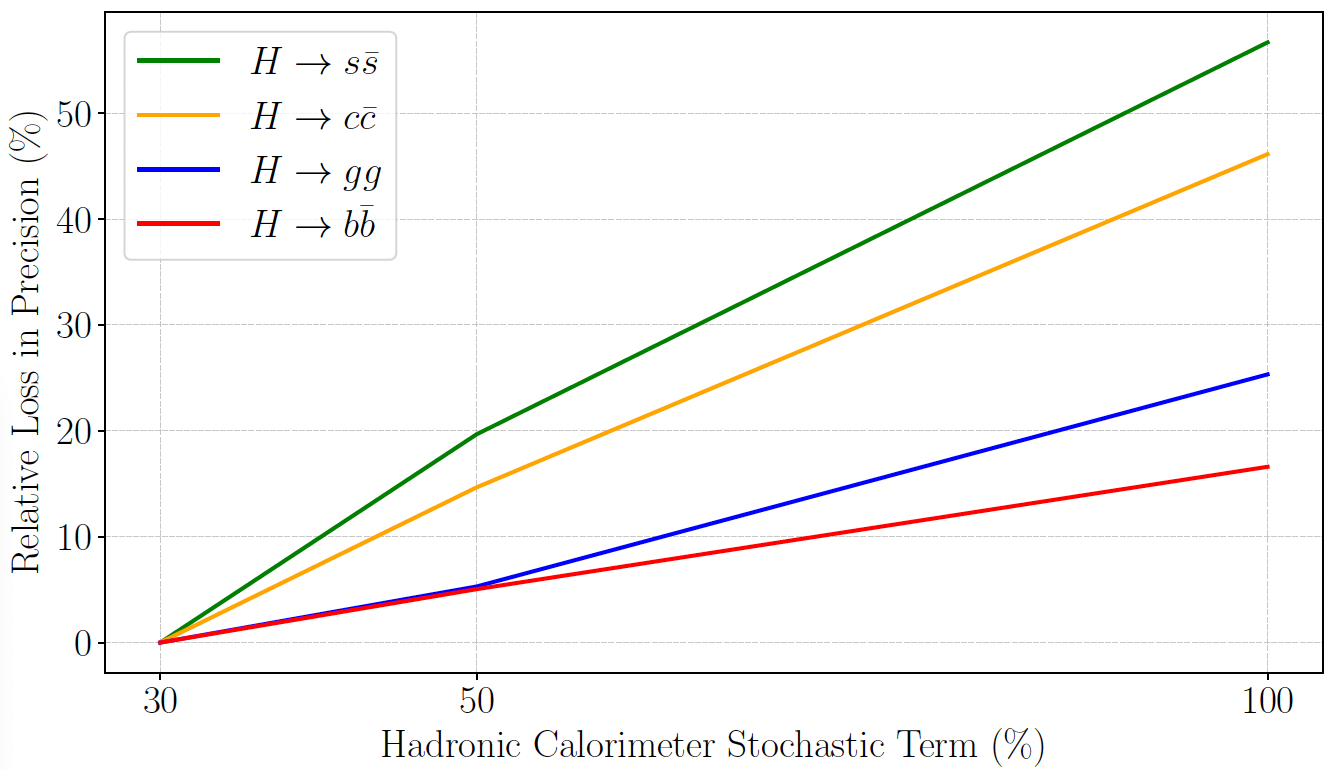"}
        \caption{Impact of HCAL resolution on Higgs coupling measurements~\cite{FCC_FS_Exp}.}
        \label{fig:hcal_higgs}
    \end{minipage}\hfill
    \begin{minipage}[b]{0.43\textwidth}
        \centering
        \includegraphics[height=0.18\textheight,trim=0cm 0cm 0cm 2cm,clip]{"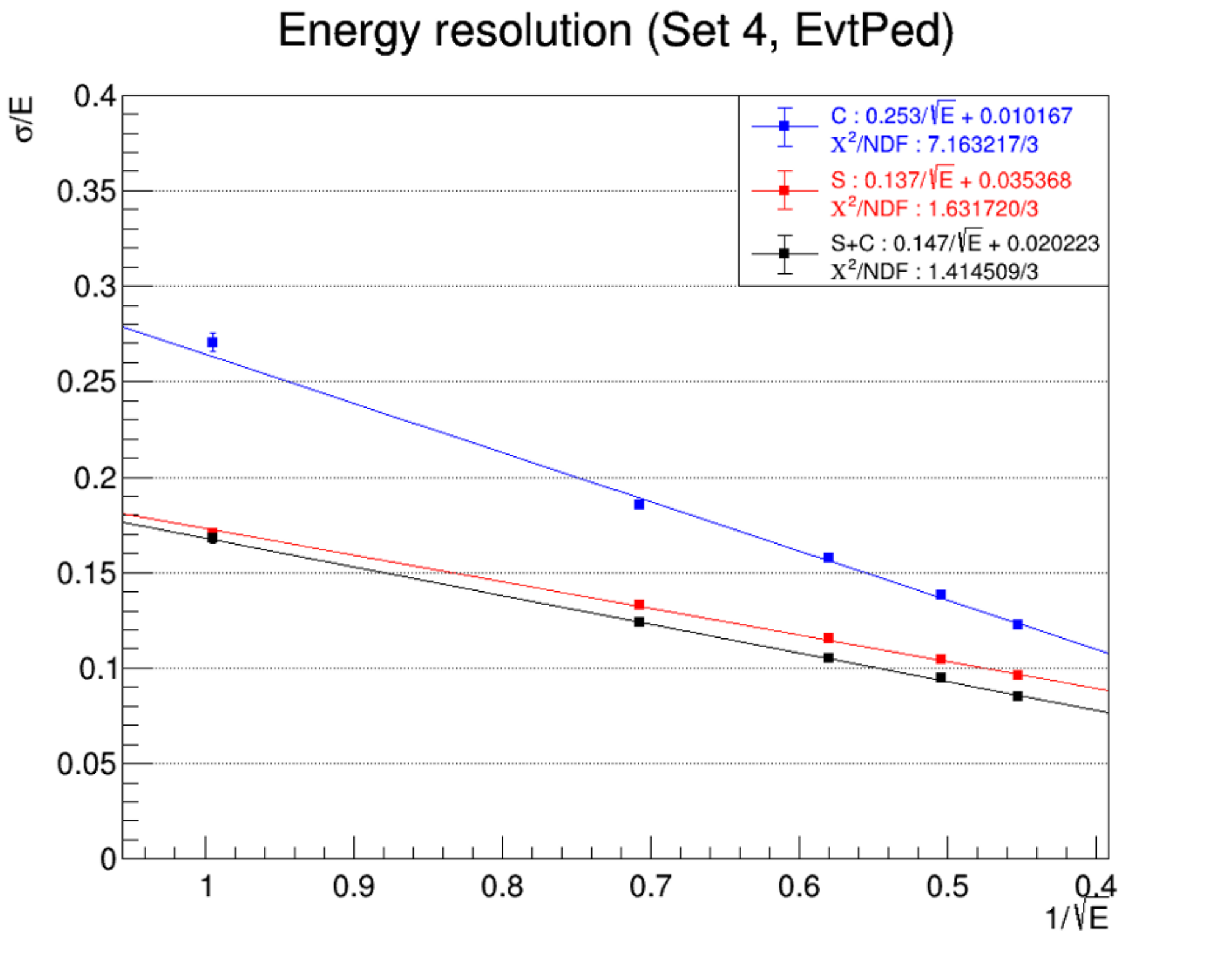"}
        \caption{Electromagnetic energy resolution of a dual-readout fibre calorimeter~\cite{Kim2025}.}
        \label{fig:hcal_resolution}
    \end{minipage}
\end{figure}

The calorimeter system is completed by the dual-readout fibre hadronic calorimeter. Scintillating fibres and clear fibres, which produce only Cherenkov light, are embedded in metal tubes that act as absorbers. These tubes are alternated to form a hexagonal pattern. Trapezoidal towers of such tubes, pointing to the IP, make up the HCAL. The goal is to reach about $\SI{30}{\percent}/\sqrt{E}$ standalone hadronic resolution, which would improve the measurement precision of the Higgs Yukawa couplings by up to $\SI{20}{\percent}$ compared to a stochastic term of only $\SI{50}{\percent}$, as shown in Figure~\ref{fig:hcal_higgs}.

Different mechanical procedures to produce the absorber structure are being studied, including 3D metal printing, square-shaped plates, and blocks of heat sink; the goal is to optimise the filling factor of absorber and fibres~\cite{Kim2025}. Figure~\ref{fig:hcal_resolution} shows the performance of a fibre HCAL prototype in a positron test beam of different energies. An electromagnetic energy resolution of $\sigma_E/E = \SI{14.7}{\percent}/\sqrt{E} \oplus \SI{2.0}{\percent}$ is achieved.

The current HCAL design features around hundred million individual tubes. This number could be optimised, given the high granularity of the crystal ECAL in front. Another research direction is the readout of the full signal spectrum (time and height) from each fibre, which would provide longitudinal segmentation and thus access to jet substructure. Lastly, clear fibres with different refractive indices could be used to lower the Cherenkov threshold for protons~\cite{IDEAStudyGroup:2025gbt}. The next step in the HCAL R\&D is the recently completed HiDRa prototype, which can contain the full hadronic shower and aims to demonstrate the scalability of the technology. HiDRa has been put into a test beam and first results are expected soon.

\section{\textmu-RWELL muon detector}

\begin{figure}[htbp]
    \centering
    \begin{minipage}[b]{0.48\textwidth}
        \centering
        \includegraphics[height=0.18\textheight]{"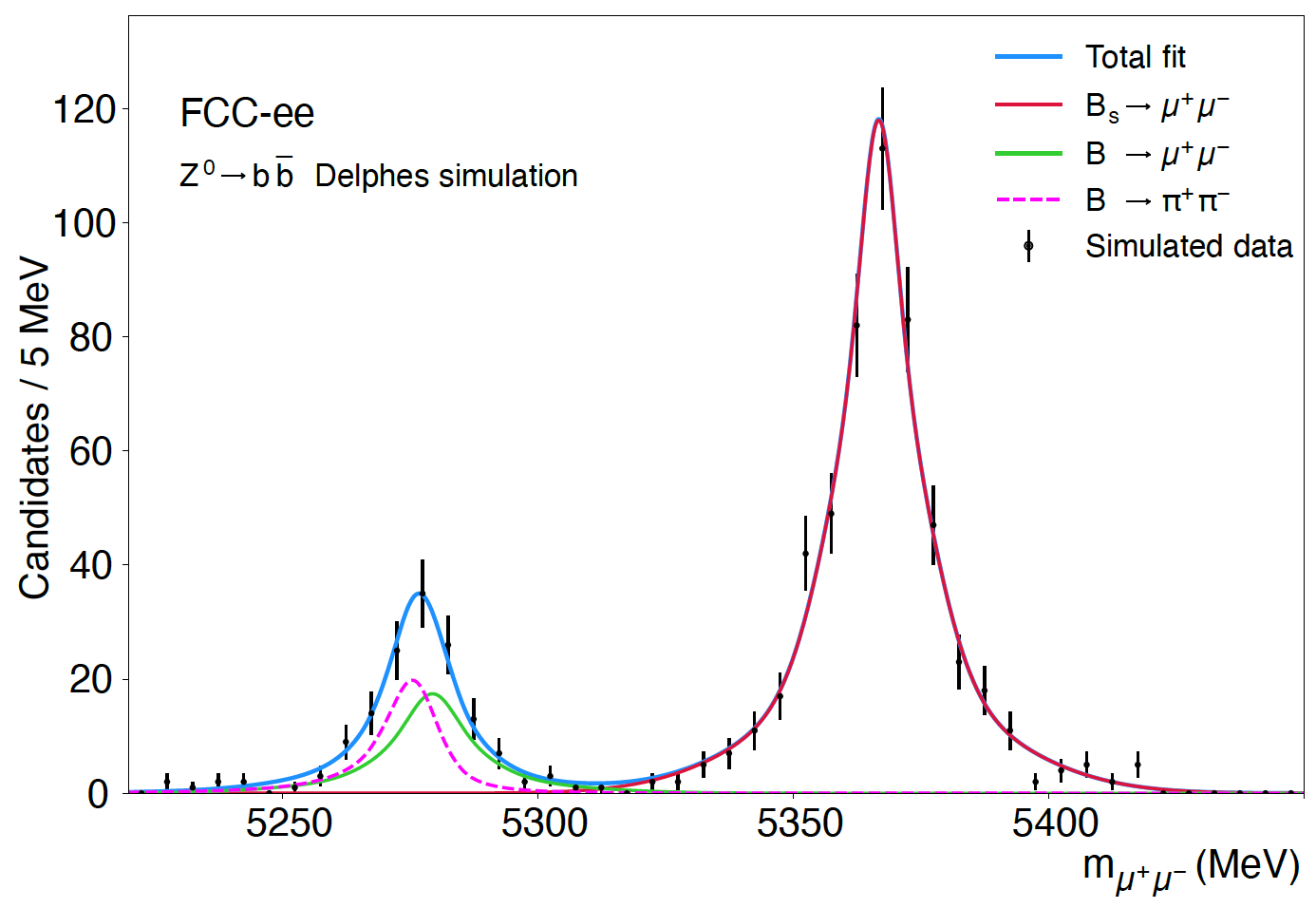"}
        \caption{Reconstructed invariant mass of $B^0$ and $B_s$, showing the background contribution of misidentified pion pairs~\cite{Monteil:2021ith}.} 
        \label{fig:muon_identification}
    \end{minipage}
    \hfill
    \begin{minipage}[b]{0.48\textwidth}
        \centering
        \includegraphics[height=0.18\textheight]{"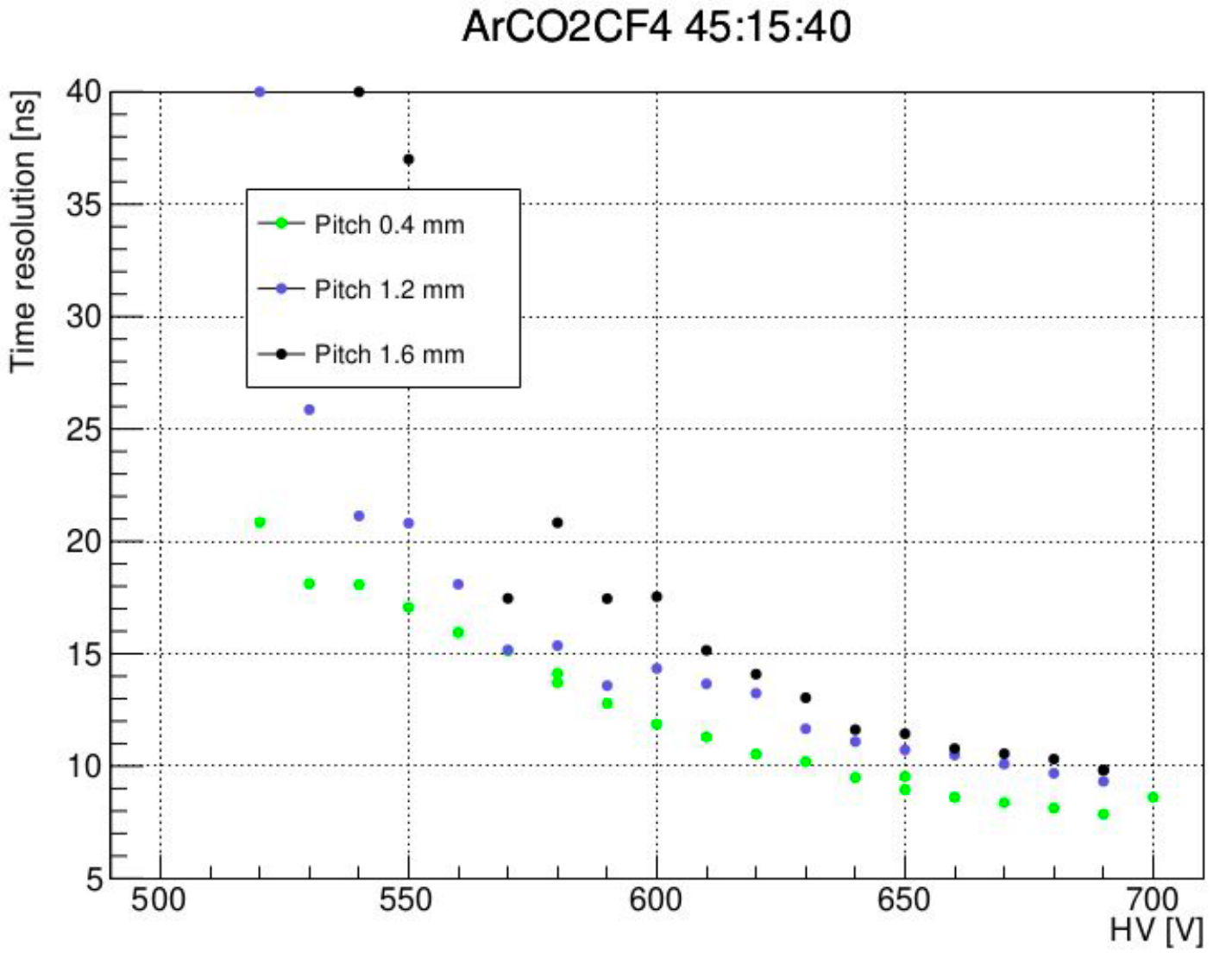"}
        \caption{Timing performance of \textmu-RWELL detector with different strip pitches, read out by the TIGER ASIC and the GEMROC FPGA.} 
        \label{fig:muon_tiger}
    \end{minipage}
\end{figure}

The track resolution, and thus momentum resolution of muons in FCC-ee experiments is driven by the tracker. A dedicated muon detector system, however, is still necessary to enable a good muon identification, efficiently rejecting for example fake pions. This is demonstrated in the measurement of $B_s \rightarrow \mu^+ \mu^-$, which would otherwise be contaminated by $B_0 \rightarrow \pi^+ \pi^-$ decays with both pions misidentified as muons, as shown in Figure~\ref{fig:muon_identification}. The muon detector is also responsible for catching the hadronic shower tails not contained in the HCAL. Furthermore, standalone particle tracking in the muon system is interesting for the search for long-lived particles decaying outside the tracker volume. The IDEA muon detector foresees at least three barrel and endcap layers of $50 \times \SI{50}{\centi\meter\squared}$ large tiles of \textmu-RWELL detectors~\cite{DiFiore:2025spq}. The tiles are overlapped to avoid dead areas.

Current R\&D compares different approaches for 2D segmentation of the sensors, either through capacity sharing anodes~\cite{Gnanvo:2023tgf}, two 1D readout schemes, or with a 1D readout with a strip-patterned top electrode for the 2D information~\cite{IDEAStudyGroup:2025gbt}. Also the front-end electronics side is being studied, with the TIGER ASIC~\cite{Amoroso:2021glb} being a promising candidate for further design development, reaching time resolutions on the order of the bunch crossing time of FCC-ee or below, as shown in Figure~\ref{fig:muon_tiger}.

\section{Conclusions}

The IDEA detector concept pushes the limit of achievable detector performance at the FCC-ee. Research both on the detector layout and on the components of the various subsystems is taking place all around the world. The full detector is now implemented in DD4hep full simulation. The next step is to evaluate the performance of the complete detector using both classic and machine-learning based reconstruction algorithms to optimise the IDEA detector concept as a whole. Detailed digitisation models of the various subsystems are being developed, which is crucial also to guide the discussion on the trigger and data acquisition system strategy.

\acknowledgments

The author is funded by the Swiss National Science Foundation 
under Grant no. 223515.

\bibliographystyle{JHEP}
\bibliography{biblio}

\end{document}